\def\oQ{\overline{Q}}
\def\oa{\overline{a}}
\def\D{\nabla}
\def\U{\mathcal{U}}
\def\oD{\overline{\nabla}}
\def\olambda{\overline{\lambda}}
\def\orho{\overline{\rho}}
\def\oG{\overline{G}}
\def\os{\overline{s}}
\def\Dc{\mathcal{D}}

\documentclass{PoS}

\title{Lattice Formulation of the N=4 D=3 Twisted Super Yang-Mills}

\ShortTitle{Lattice Formulation of the N=4 D=3 Twisted Super Yang-Mills}

\author{Alessandro D'Adda\\
        INFN sezione di Torino, and Dipartimento di Fisica Teorica, Universita 
di Torino, I-10125 Torino, Italy \\
        E-mail: \email{dadda@to.infn.it}}

\author{Issaku Kanamori\\
        Theoretical Physics Laboratory, RIKEN, Wako, 351-0198, Japan\\
        E-mail: \email{kanamori-i@riken.jp}}

\author{Noboru Kawamoto\\
        Department of Physics, Hokkaido University, Sapporo, 060-0810, Japan\\
        E-mail: \email{kawamoto@particle.sci.hokudai.ac.jp}}

\author{\speaker{Kazuhiro Nagata}%
        \\
        Department of Physics, Indiana University, Bloomington, 47405, IN, U.S.A.\\
        E-mail: \email{knagata@indiana.edu}}

\abstract
{
A lattice formulation of a three dimensional super Yang-Mills  
model with a twisted $N=4$ supersymmetry is proposed. 
The extended supersymmetry algebra 
of all 
eight supercharges is fully and exactly realized on the lattice 
with a modified ``Leibniz rule". 
The formulation we employ here is a three dimensional extension of the 
manifestly gauge covariant method which was developed in our 
previous proposal of Dirac-K\"ahler
twisted $N=2$ super Yang-Mills on a two dimensional lattice. 
The twisted $N=4$ supersymmetry algebra is geometrically realized on a 
three dimensional lattice with link supercharges and the use of
``shifted" (anti-)commutators.
A possible solution to the recent critiques on the link formulation will 
be discussed.   
}

\FullConference{The XXV International Symposium on Lattice Field Theory\\
                 July 30 - August 4 2007\\
                 Regensburg, Germany}

\begin{document}

\section{Introduction }

Formulating an exact supersymmetric model on a lattice is one of the most
challenging subjects in lattice field theory.
There has been already a number of works addressing this topic. 
Recently, it has been recognized that 
the so-called twisted version of supersymmetry (SUSY) 
plays a particularly important role
in formulating supersymmetric models on a lattice 
(See Ref. \cite{DKKN1,DKKN2,DKKN4}
and references therein.)
The crucial importance of twisted SUSY on the lattice
could be traced back to 
the intrinsic relation between twisted fermions
and Dirac-K\"ahler fermions. 
Based on this recognition, we proposed lattice formulations of
the $D=N=2$ super BF and Wess-Zumino models \cite{DKKN1} as well as
the $D=N=2$ twisted super Yang-Mills (SYM) \cite{DKKN2}
by explicitly constructing the Dirac-K\"ahler twisted $N=2$ SUSY algebra on 
a two dimensional lattice.
The main feature of our formulation is that 
the ``Leibniz rule" on the lattice can be exactly maintained
throughout the formulation, and as a result,
the lattice action is 
invariant w.r.t. all the supercharges 
associated with the twisted SUSY algebra.
It has been also recognized in \cite{DKKN2} that, 
besides the twisted $D=N=2$ algebra, 
also the Dirac-K\"ahler twisted $D=N=4$ SUSY algebra 
could 
be realized on the lattice with the lattice Leibniz rule.
Recently, we pointed out that the $D=3\ N=4$ twisted SUSY
algebra, which has eight supercharges,
can also be consistent with the lattice Leibniz rule conditions
and then we proposed an explicit construction of
the corresponding 
SYM action on the lattice \cite{DKKN4},
which is the main topic of this proceeding. 

In recent papers the authors of \cite{Bruckmann,BC} posed some critiques 
on our formulations of the noncommutative approach \cite{DKKN1} and of the link 
approach \cite{DKKN2}. A possible answer to the critique on the 
noncommutative approach\cite{DKKN1} will be given by the analysis of a 
matrix formulation of superfields\cite{ADKS}. 
Along a similar line of arguments, 
we propose a possible 
answer to the critiques in the link approach case.

\section{Discretization of $N=4$ twisted SUSY algebra in three dimensions}

We first introduce the following $N=4$ SUSY algebra
in a Euclidean three dimensional continuum spacetime. 
\begin{eqnarray}
&&\hspace{20pt}
\{Q_{\alpha i},\overline{Q}_{j\beta}\} = 2\delta_{ij}(\gamma_{\mu})_{\alpha\beta}P_{\mu}
\label{algebra} 
\end{eqnarray}
where 
the gamma matrices, $\gamma_{\mu}$,  
can be taken as Pauli matrices,
$\gamma^{\mu}(\mu=1,2,3)\equiv (\sigma^{1},\sigma^{2},\sigma^{3})$.  
$\overline{Q}_{i\alpha}$ can be taken as the complex conjugation
of $Q_{\alpha i}$, $\overline{Q}_{i\alpha}=Q^{*}_{\alpha i}
=Q^{\dagger}_{i\alpha}$ in the continuum spacetime.

The twisting procedure can be performed
by 
introducing the twisted Lorentz generator 
as a diagonal sum of the original Lorentz and the internal rotation generators.
The resulting algebra is most naturally expressed in terms of the following
Dirac-K\"ahler 
expansion of the supercharges on the basis of the gamma matrices,
\begin{eqnarray}
Q_{\alpha i} &=& (\mathbf{1}Q+\gamma_{\mu} Q_{\mu})_{\alpha i}, \hspace{30pt}
\overline{Q}_{i\alpha} \ =\ (\mathbf{1}\overline{Q}+\gamma_{\mu}\overline{Q}_{\mu})_{i\alpha},
\end{eqnarray}
where $\mathbf{1}$ represents a two-by-two unit matrix.
The coefficients of the above expansions, $(Q,\oQ_{\mu},Q_{\mu},\oQ)$,
are called twisted supercharges of $N=4$ in a three dimensional
continuum spacetime. 
After the twisting and the expansions, the original SUSY algebra (\ref{algebra})
can be expressed as,
\begin{eqnarray}
\{Q,\oQ_{\mu}\} &=& P_{\mu}, \label{D3alg1} 
\hspace{20pt}
\{Q_{\mu},\oQ_{\nu}\} \ =\ -i\epsilon_{\mu\nu\rho}P_{\rho}, 
\hspace{20pt}
\{\oQ,Q_{\mu}\} \ =\ P_{\mu}, 
\end{eqnarray}
where $\epsilon_{\mu\nu\rho}$ is the three dimensional totally anti-symmetric 
tensor with $\epsilon_{123}=+1$. 

\begin{figure}
\begin{center}
\begin{minipage}{35mm}
\begin{center}
\includegraphics[width=30mm]{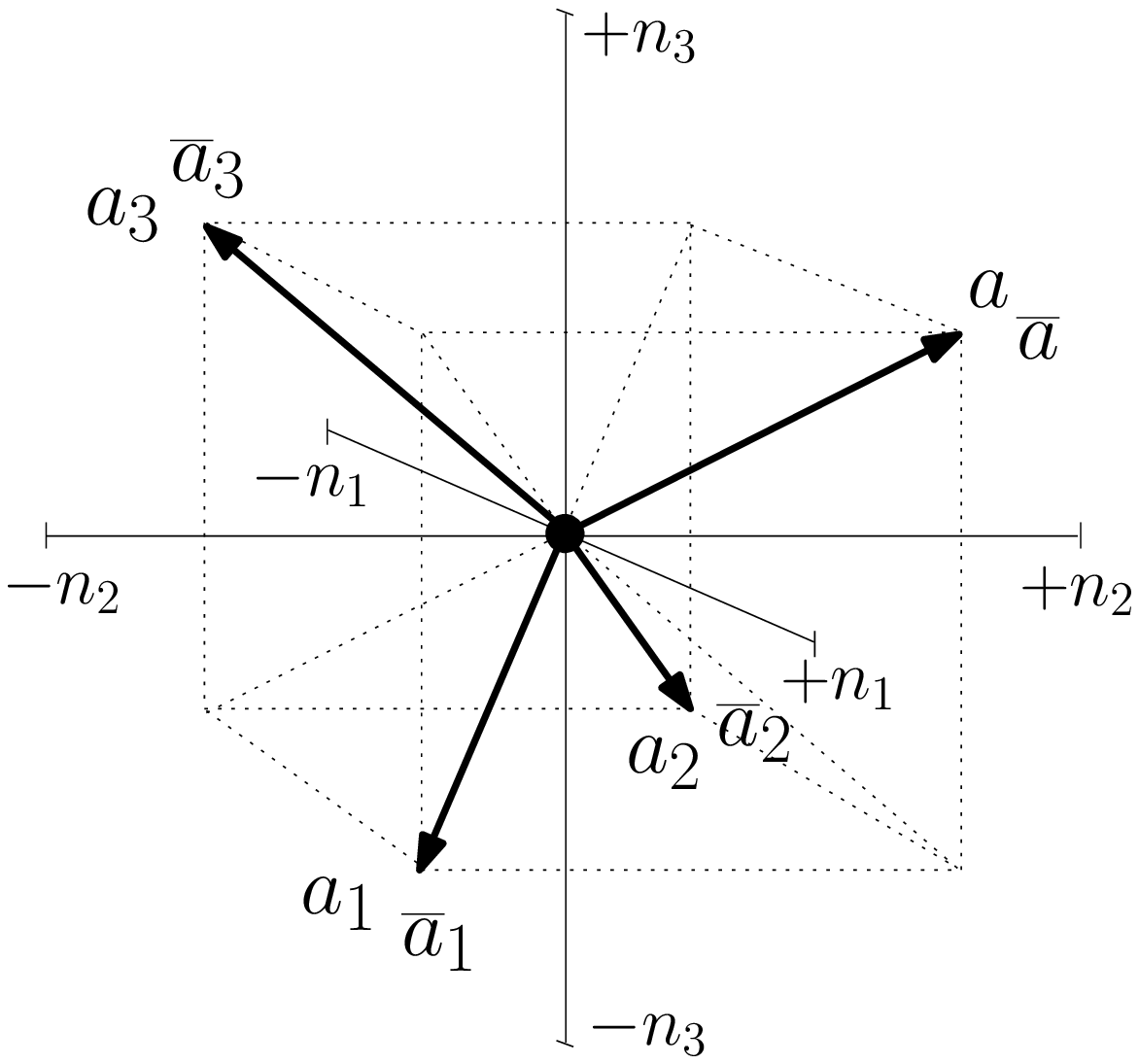}
\caption{Symmetric choice of $a_{A}$}
\label{3Dsymm_a}
\end{center}
\end{minipage}
\ \
\begin{minipage}{40mm}
\begin{center}
\includegraphics[width=40mm]{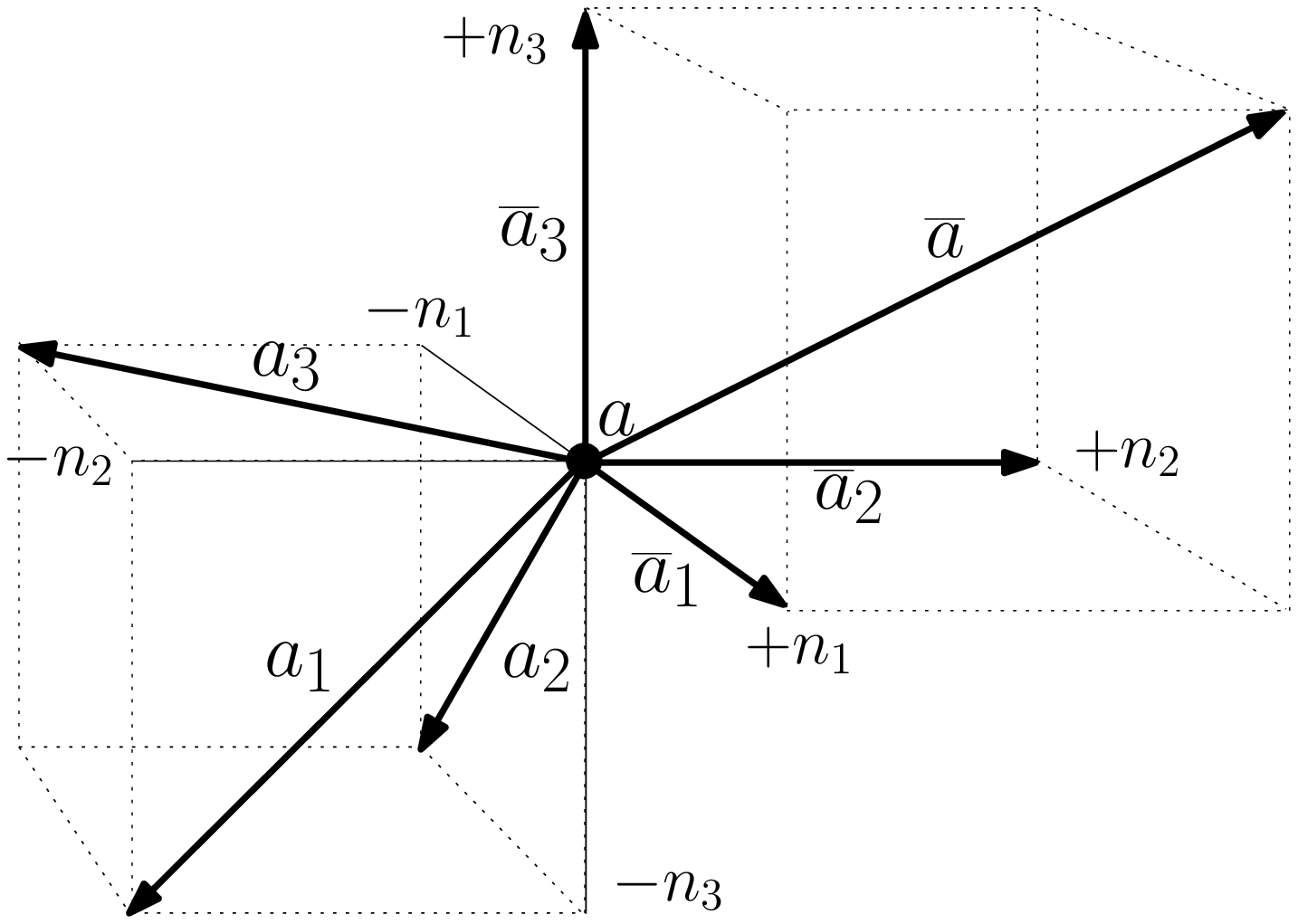}
\caption{Asymmetric choice of $a_{A}$}
\label{3Dasymm_a}
\end{center}
\end{minipage}
\ \ \ 
\begin{minipage}{70mm}
\begin{center}
\includegraphics[width=70mm]{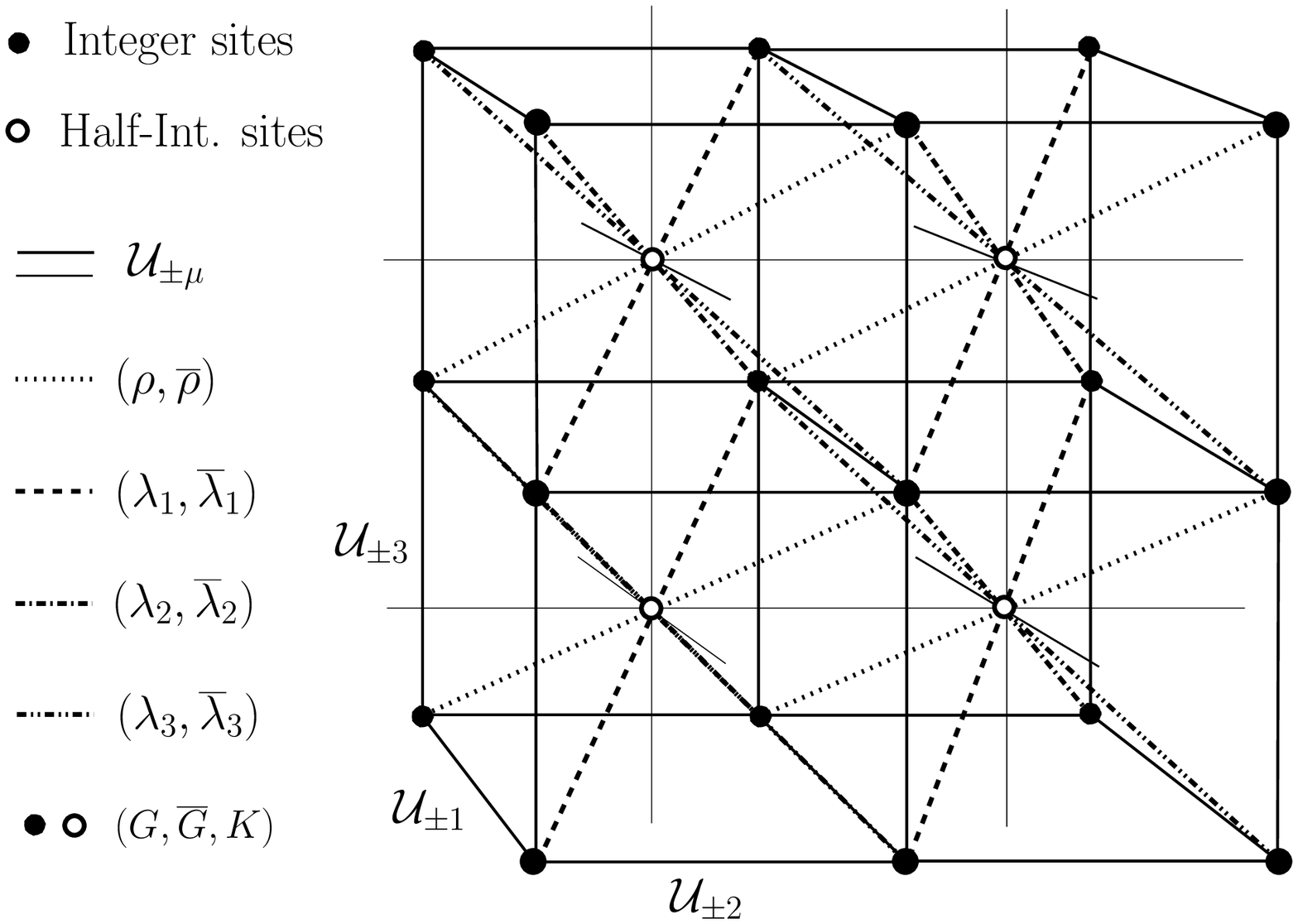}
\caption{All the configurations in the $N=4\ D=3$ twisted 
SYM action for symmetric $a_{A}$}
\label{3Dallconfig}
\end{center}
\end{minipage}
\end{center}
\end{figure}

Since we have only finite lattice spacings on a lattice, infinitesimal 
translations should be replaced by finite difference operators,
$
P_{\mu}\ =\ i\partial_{\mu} \rightarrow i\Delta_{\pm\mu},
$
where $\Delta_{\pm\mu}$ denote forward and backward difference
operators, respectively.
We locate $\Delta_{\pm\mu}$ on links from $x$ to $x\pm n_{\mu}$, respectively,
and define their operations as link commutators with shifts of $\pm n_{\mu}$.
Correspondingly, we locate the supercharges $Q_{A}$ on a link
from $x$ to $x+a_{A}$ and define its operation as a link (anti-)commutator
with a shift 
$a_{A}$.
Through these procedures, one could construct the lattice counterpart
of the SUSY algebra provided a certain type of Leibniz rule conditions hold.
It has been shown that the Dirac-K\"aher twisted type of
the $N=D=2$ and the $N=D=4$ SUSY algebra could be consistently realized 
on the lattice \cite{DKKN1,DKKN2}.
Furthermore, we recently pointed out that the $N=4\ D=3$ Dirac-K\"aher twisted 
algebra could also be formulated consistently on the lattice
and be expressed as \cite{DKKN4},
\begin{eqnarray}
\{Q,\oQ_{\mu}\} &=& +i\Delta_{+\mu}, \hspace{20pt} \label{D3latalg1} 
\{Q_{\mu},\oQ_{\nu}\} \ =\ +\epsilon_{\mu\nu\rho}\Delta_{-\rho}, 
\hspace{20pt} 
\{\oQ,Q_{\mu}\} \ =\ +i\Delta_{+\mu}, 
\end{eqnarray}
where the anti-commutators of the l.h.s are understood as link 
anti-commutators, for example,
\begin{eqnarray}
\{Q,\oQ_{\mu}\}_{x+a+\oa_{\mu},x} &=& 
Q_{x+a+\oa_{\mu},x+\oa_{\mu}}\oQ_{x+\oa_{\mu},x}
+\oQ_{x+\oa_{\mu}+a,x+a}Q_{x+a,x}. \label{link_com}
\end{eqnarray}
The corresponding Leibniz rule conditions,
\begin{eqnarray}
a+\overline{a}_{\mu} &=& +n_{\mu}, \label{leibniz1} \hspace{20pt} 
a_{\mu} + \overline{a}_{\nu} \ =\ -|\epsilon_{\mu\nu\rho}|n_{\rho}, 
\hspace{20pt} 
\overline{a}+a_{\mu} \ =\ +n_{\mu}, 
\end{eqnarray}
could be consistently satisfied by the following generic solutions,
\begin{eqnarray}
a &=& (arbitrary), \hspace{20pt}
\overline{a}_{\mu} \ =\ +n_{\mu} - a, \hspace{20pt} 
a_{\mu} \ =\ -\sum_{\lambda \neq \mu} n_{\lambda} +a, \hspace{20pt}
\overline{a} \ =\ +\sum_{\lambda =1}^{3} n_{\lambda} -a. \qquad
\end{eqnarray}
Notice that there is one vector arbitrariness 
in the choice of $a_{A}$, 
which governs the possible configurations of the three dimensional lattice.
The typical examples are the symmetric choice (Fig.\ref{3Dsymm_a}) 
and the asymmetric choice (Fig.\ref{3Dasymm_a}).
Notice that the summation of all the shift parameters
$(a,\overline{a}_{\mu},a_{\mu},\overline{a})$ vanish,
\begin{eqnarray}
\sum a_{A} &=& a + \overline{a}_{1} +\overline{a}_{2} + \overline{a}_{3}
+ a_{1} +a_{2} +a_{3} + \overline{a} \ =\ 0, 
\end{eqnarray}
regardless of any particular choice of $a_{A}$.


\section{Lattice formulation of the twisted $N=4$
SYM in three dimensions}

Based on the arguments in the previous section,
We now proceed to construct the $D=3\ N=4$ twisted 
SYM action on a Euclidean lattice.
We first introduce fermionic and bosonic gauge link variables,
$\D_{A}$ and $\U_{\pm\mu}$ which are located on links
$(x+a_{A},x)$ and $(x\pm n_{\mu},x)$, respectively,
just like $Q_{A}$ and $\Delta_{\pm \mu}$.
The gauge transformations of those link operators
are given by,
\begin{eqnarray}
(\D_{A})_{x+a_{A},x} &\rightarrow& G_{x+a_{A}}(\D_{A})_{x+a_{A},x}G^{-1}_{x}, 
\hspace{20pt} 
(\U_{\pm \mu})_{x\pm n_{\mu},x} \ \rightarrow\  
G_{x\pm n_{\mu}}(\U_{\pm \mu})_{x\pm n_{\mu},x}G^{-1}_{x},
\end{eqnarray}
where $G_{x}$ denotes the finite gauge transformation at the site $x$. 
Next we impose the following $D=3\ N=4$ twisted
SYM constraints on the lattice,
\begin{eqnarray}
\{\D,\overline{\D}_{\mu}\}_{x+a+\overline{a}_{\mu},x}
&=& +i(\U_{+\mu})_{x+n_{\mu},x}, \label{constSYM1} 
\hspace{20pt}
\{\D_{\mu},\overline{\D}_{\nu}\}_{x+a_{\mu}+\overline{a}_{\nu},x}
 \ =\ -\epsilon_{\mu\nu\rho}(\U_{-\rho})_{x-n_{\rho},x},
\\
\{\overline{\D},\D_{\mu}\}_{x+\overline{a}+a_{\mu},x} 
&=& +i(\U_{+\mu})_{x+n_{\mu},x}, \label{constSYM3} \hspace{62pt}
\{others\} \ =\ 0, \label{constSYM4}
\end{eqnarray}
where the left-hand sides should be
understood as the link anti-commutators such as in (\ref{link_com}). 

Since the multiplet of the $D=3\ N=4$ twisted SYM should contain
three components of gauge fields as well as three components of
scalar fields,
we require that the above bosonic gauge
link variables are to be defined 
in such a way to include the scalar contributions,
\begin{eqnarray}
(\U_{\pm \mu})_{x\pm n_{\mu},x} &\equiv&
(e^{\pm i(A_{\mu}\pm \phi^{(\mu)})})_{x\pm n_{\mu},x}, \label{exp}
\end{eqnarray}
where $A_{\mu}$ and $\phi^{(\mu)}(\mu =1,2,3)$ represent 
the hermitian three dimensional gauge field and the three components of the scalar
field, respectively.
Notice that the product of oppositely oriented bosonic gauge link variables
does not 
give unity, $\U_{+\mu}\U_{-\mu} \neq 1$.
It rather gives the contribution of the scalar fields.
Once imposing the SYM constraints (\ref{constSYM1})-(\ref{constSYM4}),
we automatically obtain the entire information of lattice SYM multiplet
through the analysis of
Jacobi identities.
It turns out that we have $N=4\ D=3$ twisted fermions 
$(\rho,\overline{\lambda}_{\mu},\lambda_{\mu},\overline{\rho})$
and auxiliary fields
$(G,\overline{G},K)$
besides the bosonic gauge link variables $\U_{\pm\mu}$.
All the shift properties of the component fields are summarized in 
Table \ref{shift}.
The SUSY transformation of the twisted $N=4$ lattice gauge multiplet
can be determined from the above Jacobi identity relations via
\begin{eqnarray}
(s_{A}\varphi)_{x+a_{A}+a_{\varphi},x}  \ =\ 
(s_{A})\varphi_{x+a_{\varphi},x} &\equiv&
[\D_{A},\varphi\}_{x+a_{A}+a_{\varphi},x},
\label{SUSY-tr}
\end{eqnarray} 
where $(\varphi)_{x+a_{\varphi},x}$ denotes one of the component fields
$(\U_{\pm\mu},\rho,\olambda_{\mu},\lambda_{\mu},\orho,G,\oG,K)$.
The results are summarized in Table 2.
As a natural consequence of the constraints 
(\ref{constSYM1})-(\ref{constSYM4}),
one can see that the resulting $N=4\ D=3$ twisted SUSY algebra
for the component fields closes off-shell (modulo gauge transformations) 
on the lattice.

\begin{table}
\renewcommand{\arraystretch}{1.2}
\renewcommand{\tabcolsep}{7pt}
\hfil
\begin{scriptsize}
\begin{tabular}{c|c|c|c|c|c}
\hline
& $\D$ & $\oD_{\mu}$ & $\D_{\mu}$ & $\oD$ & $\U_{\pm\mu}$  \\ \hline
shift & $a$ & $\oa_{\mu}$ & $a_{\mu}$ & $\oa$ & $\pm n_{\mu}$ 
\\ \hline
\end{tabular}
\hfil
\begin{tabular}{c|c|c|c|c|c|c|c}
\hline
& $\rho$ & $\olambda_{\mu}$ & $\lambda_{\mu}$ & $\orho$ & $G$ & $\oG$ & $K$ \\ \hline
shift & $-a$ & $-\oa_{\mu}$ & $-a_{\mu}$ & $-\oa$ & $a-\oa$ & $\oa-a$ & $0$
\\ \hline
\end{tabular}
\end{scriptsize}
\caption{Shifts carried by the link variables and the fields}
\label{shift}
\end{table}

\begin{table}
\begin{center}
\renewcommand{\arraystretch}{1.1}
\renewcommand{\tabcolsep}{10pt}
\begin{scriptsize}
\begin{tabular}{|c|c|c|c|c|}
\hline
& $s$ & $\os_{\mu}$ & $s_{\mu}$ & $\os$ \\ \hline
$\U_{+\nu}$ & $0$ & $+i\epsilon_{\mu\nu\rho}\lambda_{\rho}$
& $-i\epsilon_{\mu\nu\rho}\olambda_{\rho}$ & $0$ \\ 
$\U_{-\nu}$ & $+\olambda_{\nu}$ & $+\delta_{\mu\nu}\rho$ 
& $+\delta_{\mu\nu}\orho$ & $+\lambda_{\nu}$ \\ \hline
$\rho$ & $-K+\frac{i}{2}[\U_{+\rho},\U_{-\rho}]$ & $0$ 
& $+\frac{1}{2}\epsilon_{\mu\rho\sigma}[\U_{-\rho},\U_{-\sigma}]$ & $+\oG$ \\
$\olambda_{\nu}$ & $0$ 
& $+i[\U_{+\mu},\U_{-\nu}]$ 
& $-\delta_{\mu\nu}G$ 
& $-\frac{1}{2}\epsilon_{\nu\rho\sigma}[\U_{+\rho},\U_{+\sigma}]$
\\ 
& & $+\delta_{\mu\nu}(K-\frac{i}{2}[\U_{+\rho},\U_{-\rho}])$ & & \\
$\lambda_{\nu}$ 
& $+\frac{1}{2}\epsilon_{\nu\rho\sigma}[\U_{+\rho},\U_{+\sigma}]$ 
& $-\delta_{\mu\nu}\oG$ 
& $+i[\U_{+\mu},\U_{-\nu}]$ 
& $0$  
\\ 
& & & $-\delta_{\mu\nu}(K+\frac{i}{2}[\U_{+\rho},\U_{-\rho}])$ & \\
$\orho$ & $+G$ 
& $-\frac{1}{2}\epsilon_{\mu\rho\sigma}[\U_{-\rho},\U_{-\sigma}]$ 
& $0$ & $+K+\frac{i}{2}[\U_{+\rho},\U_{-\rho}]$ \\ \hline
$G$ & $0$ 
& $+\epsilon_{\mu\rho\sigma}[\U_{-\rho},\olambda_{\sigma}]$ 
& $0$
& $-i[\U_{+\rho},\olambda_{\rho}]$  
\\ 
& & $+i[\U_{+\mu},\orho]$ & &  \\
$\oG$ & $-i[\U_{+\rho},\lambda_{\rho}]$ & $0$ 
& $-\epsilon_{\mu\rho\sigma}[\U_{-\rho},\lambda_{\sigma}]$ & $0$ 
\\ 
& & & $+i[\U_{+\mu},\rho]$ &  \\
$K$ & $+\frac{i}{2}[\U_{+\rho},\olambda_{\rho}]$ 
& $+\frac{1}{2}\epsilon_{\mu\rho\sigma}[\U_{-\rho},\lambda_{\sigma}]$
& $+\frac{1}{2}\epsilon_{\mu\rho\sigma}[\U_{-\rho},\olambda_{\sigma}]$
& $-\frac{i}{2}[\U_{+\rho},\lambda_{\rho}]$ 
\\ 
& & $-\frac{i}{2}[\U_{+\mu},\rho]$ & $+\frac{i}{2}[\U_{+\mu},\orho]$ & \\ 
\hline 
\end{tabular}
\label{translat3DSYM}
\end{scriptsize}
\caption{SUSY trans. laws for the twisted $N=4\ D=3$ lattice SYM multiplet
$(\U_{\pm\mu},\rho,\olambda_{\mu},\lambda_{\mu},\orho,G,\oG,K)$}
\end{center}
\end{table}

The construction of the twisted $D=3\ N=4$ SUSY invariant action
can be found by noticing the ``chiral"
and ``anti-chiral" conditions of $\U_{\pm\mu}$, 
\begin{eqnarray}
S&=& +\sum_{x}\frac{1}{2}\os_{1}\os_{2}s_{1}s_{2}
\ \mathrm{tr}\ \U_{+3}\ \U_{+3} 
\ =\ -\sum_{x}\frac{1}{2}s \os_{3}\os s_{3}
\ \mathrm{tr}\ \U_{-3}\ \U_{-3}\\ 
&=& \sum_{x}\mathrm{tr}\biggl[\
\frac{1}{4}[\U_{+\mu},\U_{-\mu}]_{x,x}[\U_{+\nu},\U_{-\nu}]_{x,x} 
+K^{2}_{x,x} \nonumber \\ 
&&-\frac{1}{2}
[\U_{+\mu},\U_{+\nu}]_{x,x-n_{\mu}-n_{\nu}}
[\U_{-\mu},\U_{-\nu}]_{x-n_{\mu}-n_{\nu},x} 
+G_{x,x+\oa-a}\oG_{x+\oa-a,x}\nonumber \nonumber \\[2pt]
&&+i(\olambda_{\mu})_{x,x+\oa_{\mu}}[\U_{+\mu},\rho]_{x+\oa_{\mu},x} 
+ i(\lambda_{\mu})_{x,x+a_{\mu}}
[\U_{+\mu},\orho]_{x+a_{\mu},x} 
+\epsilon_{\mu\nu\rho}(\lambda_{\mu})_{x,x+a_{\mu}}
[\U_{-\nu},\olambda_{\rho}]_{x+a_{\mu},x}
\biggr], \nonumber \\
\label{N4D3action}
\end{eqnarray}
where the summation over $x$ should cover integer sites as well as half-integer sites 
if one takes the symmetric choice of $a_{A}$ (Fig.\ref{3Dsymm_a}),
while for the asymmetric choice (Fig.\ref{3Dasymm_a})
 it needs to cover only the integer sites.
Due to this summation property, the order in the product of 
the supercharges is shown to be irrelevant up to
total difference terms.
Notice that the exact form w.r.t. all the supercharges
and the nilpotency of each supercharge manifestly ensure
the twisted $N=4$ SUSY invariance of the action. 
It is also important to note that each term in the action
forms a closed loop, which ensures the manifest gauge invariance of the action.
This property is originated from the 
vanishing sum of the shifts associated with the action,
\begin{eqnarray}
\oa_{1}+\oa_{2}+a_{1}+a_{2}+n_{3}+n_{3} = 
a+\oa_{3}+\oa+a_{3} -n_{3}-n_{3}  = 0,
\end{eqnarray} 
which holds for any particular choice of $a_{A}$.
The gauge invariance is thus maintained
regardless of any particular choice of $a_{A}$. 
Fig.\ref{3Dallconfig} depicts all the field configurations
in 
the action (\ref{N4D3action})
in the case of the symmetric choice of $a_{A}$.

In proving the lattice SUSY invariance of the action by 
explicitly operating with the supercharge,
we need to take care about the ordering of a product of 
component fields. This is related to the critique that the authors of 
Ref.\cite{Bruckmann} pointed out that a SUSY transformation on differently 
ordered products of the same component fields gives different expressions 
and thus is inconsistent. We claim that a particular ordering of component 
fields is chosen to be the proper ordering which leads to a correct lattice 
SUSY transformation. The proper ordering of a product of component fields 
inherits the ordering of a product of the original superfields \cite{ADKS}. 
An alternative ordering of a product of component fields which leads 
to a correct lattice SUSY transformation can be obtained from the proper 
ordering by shifting the coordinates of the interchanged component fields 
as if they were noncommutative: 
\begin{eqnarray}
(\phi_A)_{x+a_A+a_B,x+a_B}(\phi_B)_{x+a_B,x}~~=~~
(-1)^{|\phi_A||\phi_B|}(\phi_B)_{x+a_A+a_B,x+a_A}(\phi_A)_{x+a_A,x}, 
\label{field-NC}
\end{eqnarray} 
where the fields $\phi_A$ and $\phi_B$ carry a shift $a_A$ and $a_B$, 
respectively. As far as the proper ordering of component fields is 
kept with the interchanging rule (\ref{field-NC}), a lattice SUSY 
transformation on a product of component fields gives a consistent 
transformation. 

Another ``inconsistency" posed in \cite{BC} is related to the link 
nature of the supercharge $s_A$ and of the supercovariant derivative $\D_A$. 
A SUSY transformation $s_A$ on the action generates a link hole 
$(x+a_A,x)$ since all the terms in the action have a vanishing shift 
and thus are composed of closed loops. At first look a naive 
supercharge operation to the action leads to gauge variant terms since 
such terms have link holes. We claim that we need to introduce 
a covariantly constant fermionic parameter $\eta_B$ which anti-commutes 
with all supercovariant derivatives in the shifted anti-commutator sense, 
\begin{eqnarray}
\{\D_A,\eta_B\}_{x+a_A-a_B,x} 
&=& (\D_A)_{x+a_A-a_B,x-a_B}(\eta_B)_{x-a_B,x} + 
(\eta_B)_{x+a_A-a_B,x+a_A}(\D_A)_{x+a_A,x} 
\ =\ 0, \qquad
\label{fermionic-P}
\end{eqnarray} 
where $\eta_B$ has a shift $-a_B$ and thus can fill up the link holes 
to generate gauge invariant terms. 
We define the gauge transformation of the superparameter, 
\begin{eqnarray}
(\eta_A)_{x-a_A,x} \rightarrow G_{x-a_A}(\eta_A)_{x-a_A,x}G_x^{-1}.
\label{gauge-tr-FP}
\end{eqnarray} 
We can then prove the exact SUSY invariance of the action by applying a 
shiftless combination of SUSY transformation $\eta_A s_A$ (no sum) to the 
action. 
The SUSY transformation of the component fields 
including this fermionic parameter is given by 
\begin{eqnarray}
(\eta_A s_{A}\varphi)_{x+a_{\varphi},x} \ = \
(\eta_A)_{x+a_{\varphi},x+a_{\varphi}+a_A} 
(s_{A}\varphi)_{x+a_{\varphi}+a_A,x},
\label{SUSY-tr-CF}
\end{eqnarray} 
where the SUSY transformation $(s_{A}\varphi)_{x+a_{\varphi}+a_A,x}$ 
is defined by (\ref{SUSY-tr}) and is given in Table 2.

The na\"ive continuum limit of the action (\ref{N4D3action})
can be taken through the expansion of the gauge link variables 
(\ref{exp}) around unity.
After using trace properties,
one obtains the following continuum action,
\begin{eqnarray}
S\rightarrow S_{cont}
&=& \int d^{3}x\ \mathrm{tr}\ \biggl[
\frac{1}{2}
F_{\mu\nu}F_{\mu\nu}
+K^{2} + G\oG \nonumber \\
&& -[\Dc_{\mu},\phi^{(\nu)}][\Dc_{\mu},\phi^{(\nu)}]
-\frac{1}{2}[\phi^{(\mu)},\phi^{(\nu)}][\phi^{(\mu)},\phi^{(\nu)}] \nonumber \\[6pt]
&&-i\olambda_{\mu}[\Dc_{\mu},\rho] - i\lambda_{\mu}[\Dc_{\mu},\orho]
+\epsilon_{\mu\nu\rho}\lambda_{\mu}[\Dc_{\nu},\olambda_{\rho}]\nonumber \\[4pt]
&& -\olambda_{\mu}[\phi^{(\mu)},\rho] - \lambda_{\mu}[\phi^{(\mu)},\orho]
+i\epsilon_{\mu\nu\rho}\lambda_{\mu}[\phi^{(\nu)},\olambda_{\rho}]
\biggr], \label{3DSYMaction_cont2}
\end{eqnarray}
where $F_{\mu\nu}\equiv i[\Dc_{\mu},\Dc_{\nu}]$ represents the field strength
with $\Dc_{\mu} \equiv \partial_{\mu} -iA_{\mu}$, while $\phi^{(\mu)}(\mu=1,2,3)$
denote the three independent hermitian scalar fields
in the twisted $N=4\ D=3$ SYM multiplet in the continuum spacetime. 
One could see that the kinetic term and the potential term as well as 
the Yukawa
coupling terms for the scalar fields naturally come up from the
contributions of zero-area loops in the lattice action. 
The above action 
(\ref{3DSYMaction_cont2}) is in complete agreement with
the continuum construction of the $N=4$ twisted SYM in three dimensions.

\section{Discussions} 

A fully exact SUSY invariant formulation of the
twisted $N=4$ SYM action on a three dimensional lattice 
is presented. 
Twisted $N=4$ SUSY invariance is a natural consequence of the exact 
form of the action with respect to all the twisted supercharges 
up to surface terms.
Possible answers to the critiques on the formulation of the link 
approach are given. 
It is pointed out that there is a proper ordering of a product 
of component fields which leads to the correct lattice SUSY transformation. 
We need to introduce superparameters which anti-commute with all the 
supercovariant derivatives. 
It would be important to find an explicit representation of such superparameters. 
We further have to accept that the structure behind the nature
of the component fields which carry a shift and satisfy the relation 
(\ref{field-NC}) still remains to be better clarified.
We consider that the lattice SUSY transformation can be defined only 
semilocally due to the next neighboring ambiguity of the difference operation 
and thus 
affects the ordering of component fields. 
Superfields may be able to take care of this semilocal nature of SUSY 
transformation faithfully\cite{ADKS}.  

Although we have not addressed the issue of hermiticity in detail, 
it is possible 
to understand 
hermiticity properties and Majorana nature of fermions 
in the two dimensional formulation\cite{DKKN3}.
We recognize that hermiticity properties of lattice SYM should 
be clarified
through a better geometrical 
understandings of chirality on the lattice.
It should also be mentioned that a dimensional reduction of 
the three dimensional $N=4$ twisted SYM
could give us a formulation of the $N=4$ twisted SYM 
on a two dimensional lattice, which corresponds to a double charged
system of the $N=D=2$ twisted SYM\cite{DKKN3}.
It is also important to proceed to perform a possible lattice formulation of 
the $N=D=4$  Dirac-K\"ahler twisted SYM 
which should be carried out basically in the same manner as
presented here.     
The results of these analyses will be given elsewhere.

\subsection*{Acknowledgments}

We would like to thank to J.~Kato, A.~Miyake and J.~Saito for useful discussions.
This work is supported in part by the Japanese Ministry of Education,
Science, Sports and Culture under the grant number 18540245 and  
also by INFN research funds.
I.K. is supported by the Special Postdoctoral Researchers Program at RIKEN.
K.N. is supported by Department of Energy US Government, 
Grant No. FG02-91ER 40661.


\begin{thebibliography}{99}

\bibitem{DKKN1}
  A.~D'Adda, I.~Kanamori, N.~Kawamoto and K.~Nagata,
  Nucl.\ Phys.\ {\bf B707} (2005) 100
  [{\tt hep-lat/0406029}],
  Nucl. Phys. Proc. Suppl. {\bf 140} (2005) 754
	[\texttt{hep-lat/0409092}],
  Nucl. Phys. Proc. Suppl. {\bf 140} (2005) 757.

\bibitem{DKKN2}
  A.~D'Adda, I.~Kanamori, N.~Kawamoto and K.~Nagata,
  Phys.\ Lett.\ {\bf B633} (2006) 645
  [{\tt hep-lat/0507029}]. 

\bibitem{DKKN4}
  A.~D'Adda, I.~Kanamori, N.~Kawamoto and K.~Nagata,
 {\tt arXiv:0707.3533 [hep-lat]}.

\bibitem{Bruckmann}

F.~Bruckmann and M.~de Kok,
Phys.\ Rev.\  D {\bf 73}, 074511 (2006)
[{\tt hep-lat/0603003}].

\bibitem{BC}

F.~Bruckmann, S.~Catterall and M.~de Kok,
Phys.\ Rev.\  D {\bf 75}, 045016 (2007)
[{\tt hep-lat/0611001}].

\bibitem{ADKS}

S.~Arianos, A.~D'Adda, N.~Kawamoto and J.~Saito, to appear. 


\bibitem{DKKN3}
 A. D'Adda, I. Kanamori, N. Kawamoto, K. Nagata and J. Saito, 
 to appear.


\end{thebibliography}
\end{document}